\documentclass[aps,twocolumn,showpacs,floatfix,superscriptaddress,preprintnumbers]{revtex4-1}
\usepackage{amsfonts}
\usepackage{amssymb}
\usepackage{amsmath,mathtools}
\usepackage{subfigure}

\RequirePackage{ifpdf}
\ifpdf
  \usepackage[pdftex]{graphicx}
\else
  \usepackage[dvipdfmx]{graphicx}
\fi

\usepackage{url}

\renewcommand{\vec}[1]{\mathbf{#1}}

\usepackage{color}

\begin{document}
\preprint{OU-HET-1031}

\title{Escape from black holes in materials: \\
Type II Weyl semimetals and generic edge states}
\author{Koji Hashimoto}
\email{koji@phys.sci.osaka-u.ac.jp}
\noaffiliation
\author{Yoshinori Matsuo}
\email{matsuo@het.phys.sci.osaka-u.ac.jp}
\noaffiliation
\affiliation{Department of Physics, Osaka University, Toyonaka, Osaka 560-0043, Japan}

\begin{abstract}
Type II Weyl semimetals are
dictated by bulk excitations with tilted light cones, resembling the inside of black holes.
We obtain generic boundary conditions for surface boundaries of the type II Weyl semimetals near Weyl nodes, 
and show that
for a certain boundary condition edge states can escape out of the ``black hole" event horizon.
This means that for realization of the material ``black hole" by the Type II Weyl semimetals 
a careful choice of the boundary condition is necessary.
\end{abstract}


\maketitle

\setcounter{footnote}{0}

\noindent

\section{Introduction}

Among various interplay between condensed matter physics and particle physics, recent 
advances in physics on Weyl semimetals (see \cite{armitage2018weyl} for a recent review) is of particular interest, 
because of its uniqueness about relativistic nature of quasiparticle excitations.
Study of Weyl fermions in the Weyl semimetals enlarges the common grounds of the two subjects,
not only through the anomaly and topological nature of Weyl fermions leading to the
bulk-edge correspondence \cite{jackiw1976solitons,hatsugai1993chern,wen2004quantum}, 
but also with relativistic properties of Weyl fermions.

An intriguing picture of the latter was proposed by Volovik and Zhang \cite{volovik2017lifshitz}, 
concerning in particular  
Type II Weyl semimetals \cite{soluyanov2015type}. Type II Weyl semimetals are defined
by Weyl points associated with overtilted Weyl cones, and Ref.~\cite{volovik2017lifshitz}
clarified that they correspond to light cones allowing propagation only in a 
certain direction, which in particle physics typically appears behind event horizons of black holes.

In this paper we combine theoretically the idea \cite{volovik2017lifshitz} of equivalence between the
Type II Weyl semimetals and black holes, and the
bulk-edge correspondence. We analyze most generic edge dispersion of continuum 
Type II Weyl semimetals.
The aim is to study whether the idea of identifying the Type II Weyl semimetals with the inside of the 
black holes is valid even with the presence of the edge modes.
We follow the strategy developed in Ref.~\cite{hashimoto2017boundary} 
on all possible allowed boundary conditions in the continuum limit to seek for a possibility of escaping out of the ``black hole."
We find that for a certain class of the 
boundary conditions of the surface of the semimetal, edge modes can escape from
the black hole. This means that the identification needs a proper choice of the boundary condition.

The organization of this paper is as follows. First, in Section \ref{sec2}
we briefly review continuum Type II Weyl semimetals and their relation to
black holes. Then in Section \ref{sec3} we introduce generic boundary condition analysis for
Type II Weyl semimetals with surfaces.
In Section \ref{sec4} we explicitly calculate the generic edge dispersion of Type II Weyl semimetals.
In Section \ref{sec5} we provide a useful theorem that any edge dispersion is tangential to
and ending at bulk dispersion, for generic Weyl semimetals.
Then finally in Section \ref{sec6} we calculate spacetime light cone structure for the edge modes
and find that they can escape from the black hole for a choice of the surface boundary conditions.
The final section is for a summary and discussions. 

\begin{figure*}
\begin{center}
\includegraphics[scale=0.4]{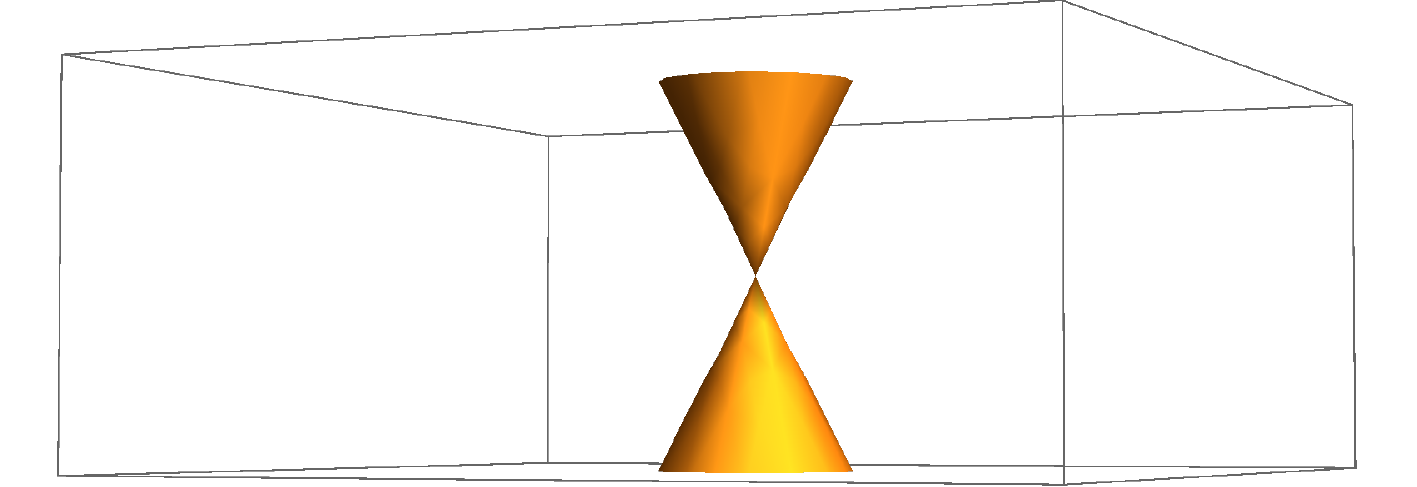}
\includegraphics[scale=0.4]{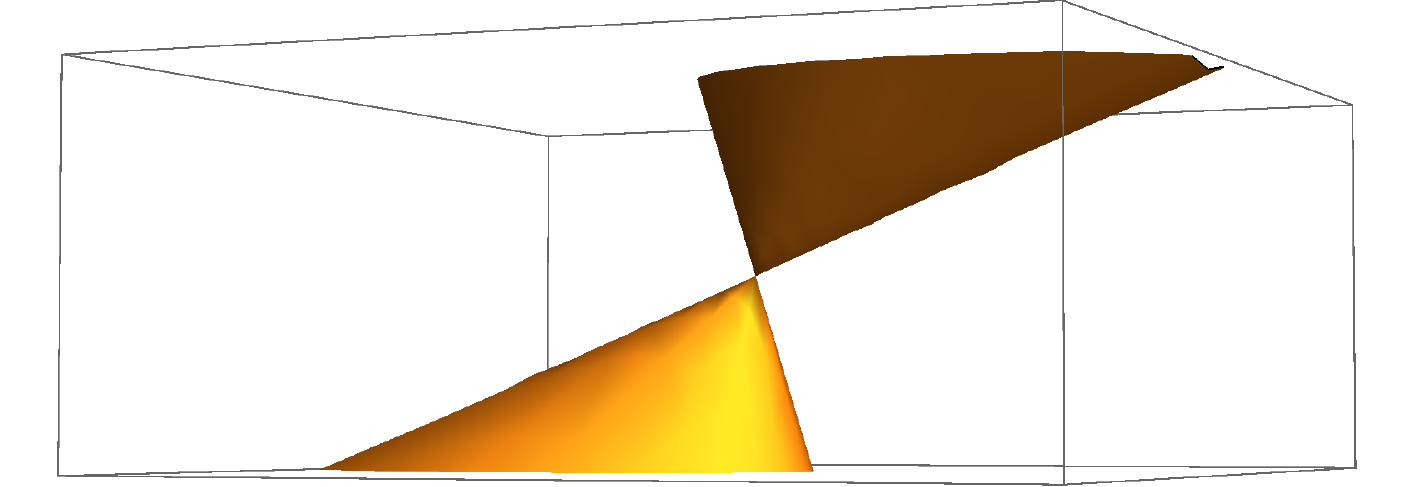}
\includegraphics[scale=0.4]{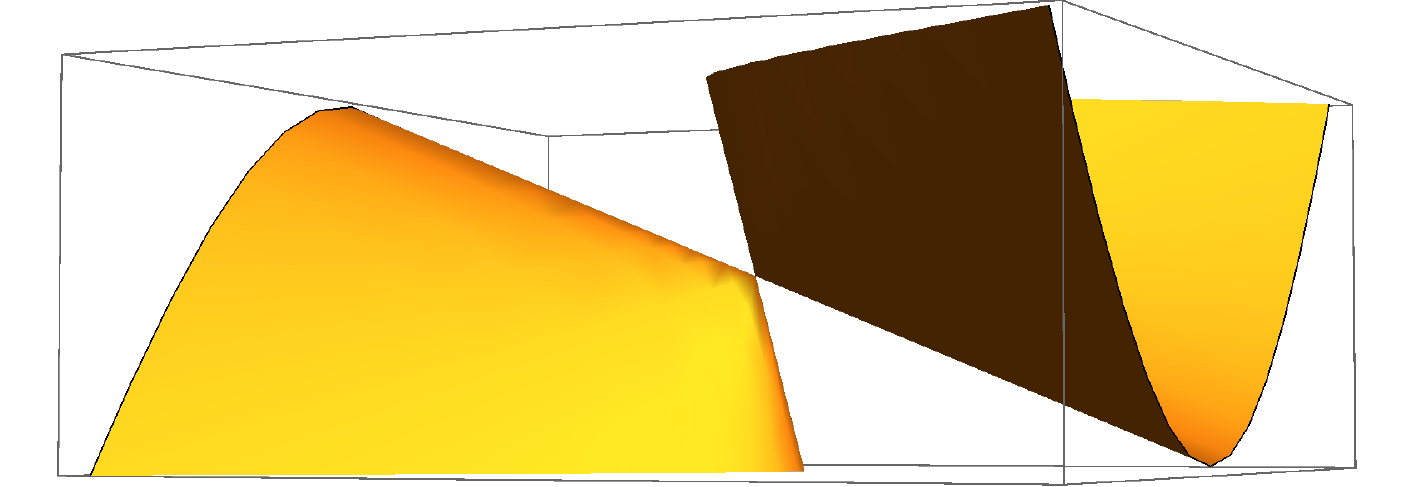}
\\[3mm]
\includegraphics[scale=0.4]{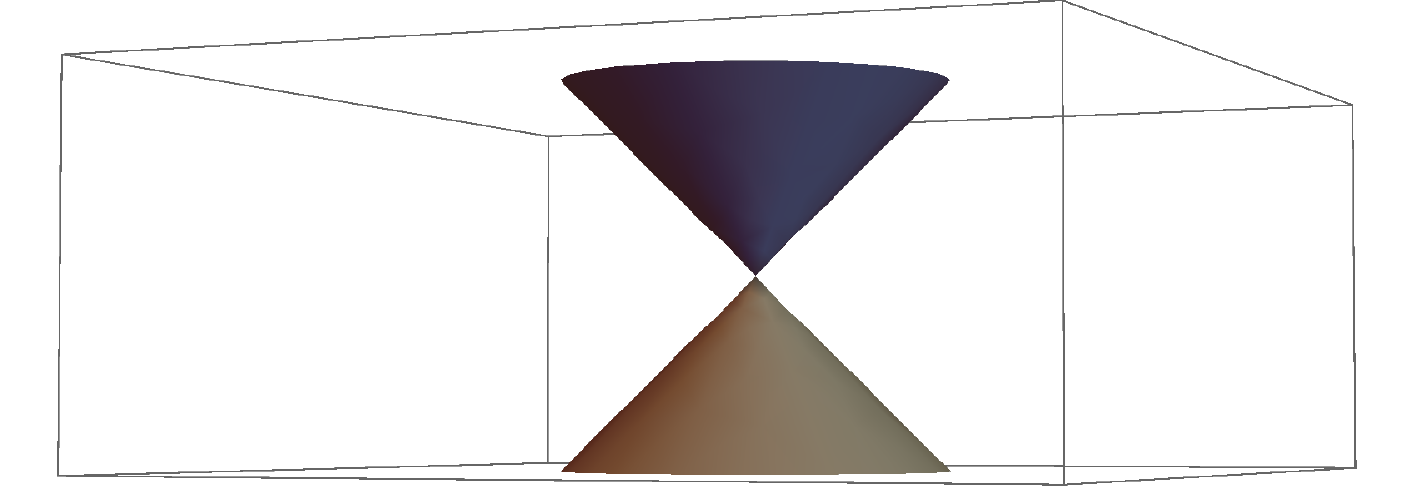}
\includegraphics[scale=0.4]{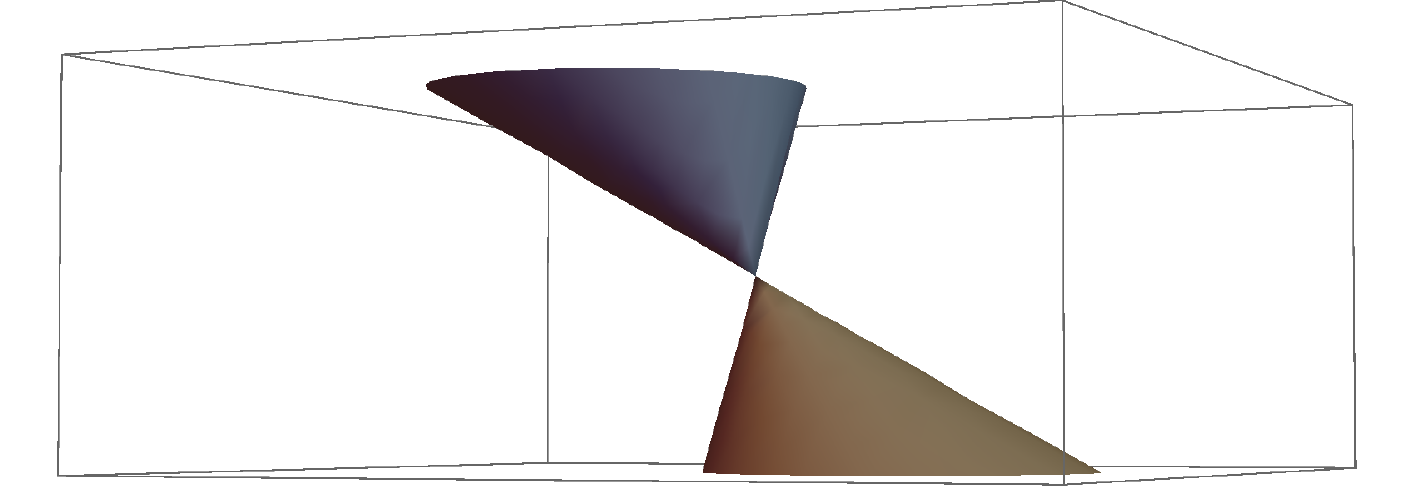}
\includegraphics[scale=0.4]{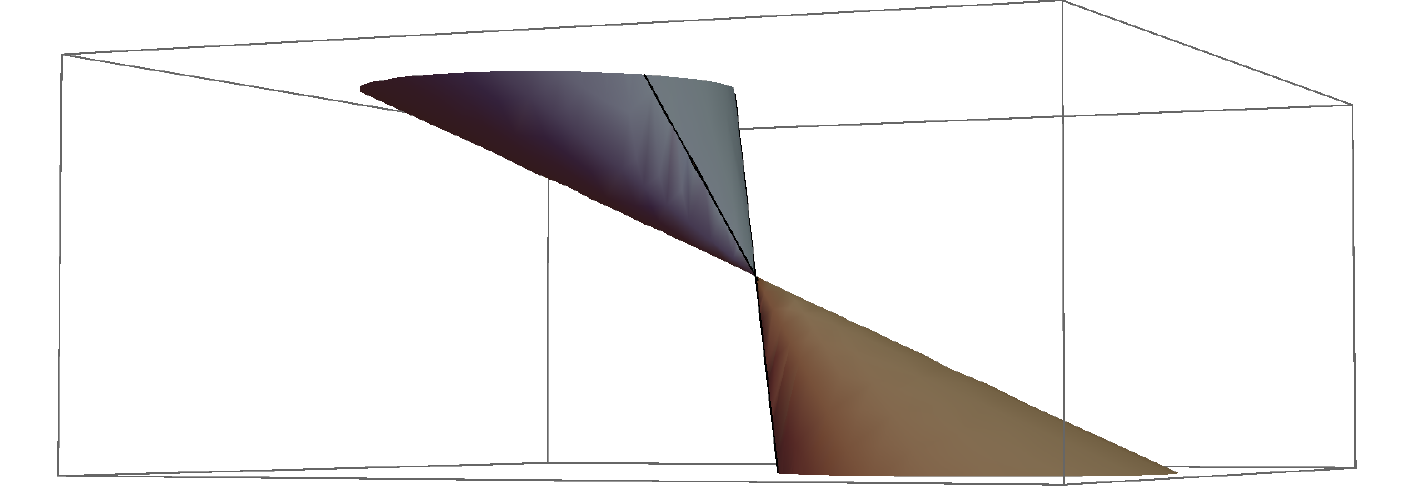}
\end{center}
\vspace{-3mm}
\caption{
Upper row: Energy dispersion $E_{\rm bulk}$ as a function of $p_1$ and $p_2$ at the slice $p_3=0$.
For simplicity we choose $\alpha_2=\alpha_3=0$, with $\alpha_1=0$ (Left, Type I), $\alpha_1 = -0.8$
 (Middle, Type I), $\alpha_1=-1.2$ (Right, Type II), respectively. 
 Lower row: corresponding light cones.
It is seen that the Type II dispersion (Right) has a large tilt of the light cone
such that it allows only a propagation to the negative direction of $x^1$.
}
\label{Lightconefig}
\end{figure*}



\section{Type II Weyl semimetals and black holes}
\label{sec2}

Let us briefly review the relation between the Type II Weyl semimetals
and light cone structure \cite{volovik2017lifshitz, volovik2016black}. 
We consider a 3-dimensional Weyl semimetal in the continuum limit,
whose Hamiltonian is given by
\begin{align}
H = p_i \sigma_i + \alpha_i p_i \mathbf{1}
\label{Hamiltonian}
\end{align}
where the summation is made for $i=1,2,3$ and $\sigma_i$ is the Pauli matrices.
This Hamiltonian is general enough to capture the topological 
charge of the Weyl semimetal, chirality $=+1$, after a proper redefinition of
the momentum axis and its normalization. The parameters 
$\alpha_i\,  (i=1,2,3)$ are real constants.
\footnote{Since the Hamiltonian \eqref{Hamiltonian} is the low energy approximation,
in reality there exists higher order terms in momenta. However, inclusion of those
higher order terms will spoil the spacetime interpretation presented here,
as those are not effectively described by the emergent metric in general.}

The bulk dispersion which follows from \eqref{Hamiltonian} is
\begin{align}
E_{\rm bulk} =
\alpha_i p_i \pm |\vec{p}|.
\label{Ebulk}
\end{align}
For $(\alpha_i)^2>1$, the bulk dispersion at $E=0$ is not a single point,
but forms a set of flat surfaces in the momentum space, 
which defines the Type II Weyl semimetals. 
See fig.~\ref{Lightconefig}.

Let us derive the light cone structure of the propagation of the excitation from the 
dispersion relation \eqref{Ebulk}. It can be recast to the form
$g^{\mu\nu}p_\mu p_\nu = 0$ with the effective metric
\begin{align}
g_{\mu\nu} = \left(
\begin{array}{cccc}
1-\alpha_i^2 & \alpha_1 & \alpha_2 & \alpha_3 \\
\alpha_1 & -1 & 0 & 0\\
\alpha_2 & 0 & -1 & 0 \\
\alpha_3 & 0 & 0 & -1
\end{array}
\right)_{\mu\nu}
\label{effmet}
\end{align}
with the standard identification $p_0 = -E$ (to make sure that the wave function is written as
$\exp[-i E t + i p_i x^i]$).

In the following we show in two ways that this is the metric inside of a black hole. For simplicity we
consider $\alpha_2=\alpha_3 = 0$. First, consider a Schwarzschild black hole metric in 
Painlev\'e-Gullstrand coordinates, 
\begin{align}
ds^2 = \left(1\!-\!\frac{2M}{r}\right)dt^2 - 2\sqrt{\frac{2M}{r}}dt dr - dr^2 - r^2 d\Omega_2^2.
\end{align}
Expand the metric around a spatial point near the 
horizon $(x,y,z)=(2M+\delta x,0,0)$ and denote a coordinate $\alpha^2 \equiv 2M/(2M+\delta x)$,
then
\begin{align}
ds^2 = (1-\alpha^2) dt^2 + 2\alpha dt dx -dx^2- dy^2-dz^2.
\end{align}
This reproduces the effective metric of the Weyl semimetal, \eqref{effmet}. 
If the expansion point is inside of the black hole, $\delta x < 0$,
then $\alpha^2>1$, so it corresponds to the dispersion of  the Type II Weyl semimetal.

Another way to see a relation to the black hole is an explicit construction of light cones. 
A null vector $n^\mu$ satisfies $n^\mu n^\nu g_{\mu\nu}=0$, which is
\begin{align}
\frac{1}{\alpha^2-1}(n^x)^2 = (\alpha^2-1)(\tilde{n}^t)^2 + (n^y)^2 + (n^z)^2
\end{align}
with $\tilde{n}^t \equiv n^t + \frac{\alpha}{1-\alpha^2}n^x$. The section at $n^y=n^z=0$
is given by
\begin{align}
\frac{n^t}{n^x} = \frac{\alpha\pm 1}{\alpha^2-1}
\end{align}
which is always negative (positive) for $\alpha < -1$ ($\alpha > 1$).
This means that the light propagation is always in a certain direction, it never goes back,
which happens also 
inside a ``black hole."
See Fig.~\ref{Lightconefig} for a pictorial view of the light cone structure.

\section{Generic boundary conditions for Type II Weyl semimetals}
\label{sec3}

Following Ref.~\cite{hashimoto2017boundary}, here we obtain the most generic boundary 
conditions for the Type II Weyl semimetals in the continuum limit.\footnote{See also Refs.~\cite{tanhayi2016role,kharitonov2017universality,candido2018paradoxical} for 1d and 2d generic
boundary conditions.}

For the Type II Weyl material, 
we introduce a single flat boundary surface at $x^3=0$, with a generic boundary condition
\begin{align}
N \psi(x^3=0) = 0
\label{N}
\end{align}
where $N$ is a constant complex $2\times 2$ matrix. With the Hamiltonian \eqref{Hamiltonian}, 
the hermiticity condition for the system
requires
\begin{align}
\psi_1^\dagger \left(\sigma_3 + \alpha_3 \mathbf{1}\right)\psi_2 = 0
\label{hermiticity}
\end{align}
for arbitrary wave functions $\psi_1$ and $\psi_2$ at the boundary.
We like to find the most generic $N$ which leads to \eqref{hermiticity}.
First, noting $\det N=0$ from \eqref{N}, we can write $N$ as
\begin{align}
N = \left(
\begin{array}{cc}
1 & \beta \\ \gamma & \gamma \beta
\end{array}
\right), 
\end{align}
up to the overall normalization of $N$ (which is irrelevant to the boundary condition \eqref{N}), 
so the solution of \eqref{N} is written as $\psi_i = (-\beta, 1)^{T} f_i$
with a scalar function $f_i$. Then the condition \eqref{hermiticity}
is recast to
\begin{align}
|\beta|^2-1 + \alpha_3(|\beta|^2+1)=0.
\end{align}
So we find that a consistent boundary condition exists only when $|\alpha_3|<1$ and  
$|\beta| = \sqrt{\frac{1-\alpha_3}{1+\alpha_3}}$. In other words, the most general 
boundary condition for Weyl semimetals with the Hamiltonian \eqref{Hamiltonian} is
\begin{align}
\left(
1,\sqrt{\frac{1-\alpha_3}{1+\alpha_3}} e^{i\theta}
\right) \psi(x^3=0) = 0
\label{bc}
\end{align}
with a boundary condition parameter $\theta$ ($0\leq \theta < 2\pi$).

Note that introduction of the boundary at $x^3=0$ does not allow $|\alpha_3|>1$. 
This also implies that the vector $\mathbb \alpha$ of Type II Weyl semimetals cannot be normal to the boundary.\footnote{
This bound was  independently studied in Ref.~\cite{zyuzin2018flat}.
The authors would like to thank A.~Zyuzin for bringing Ref.~\cite{zyuzin2018flat} to our attention.
}
Of course, putting $\alpha_3=0$ brings us back to the generic boundary condition
studied in Ref.~\cite{hashimoto2017boundary}.

\begin{figure*}
\begin{center}
\includegraphics[scale=0.3]{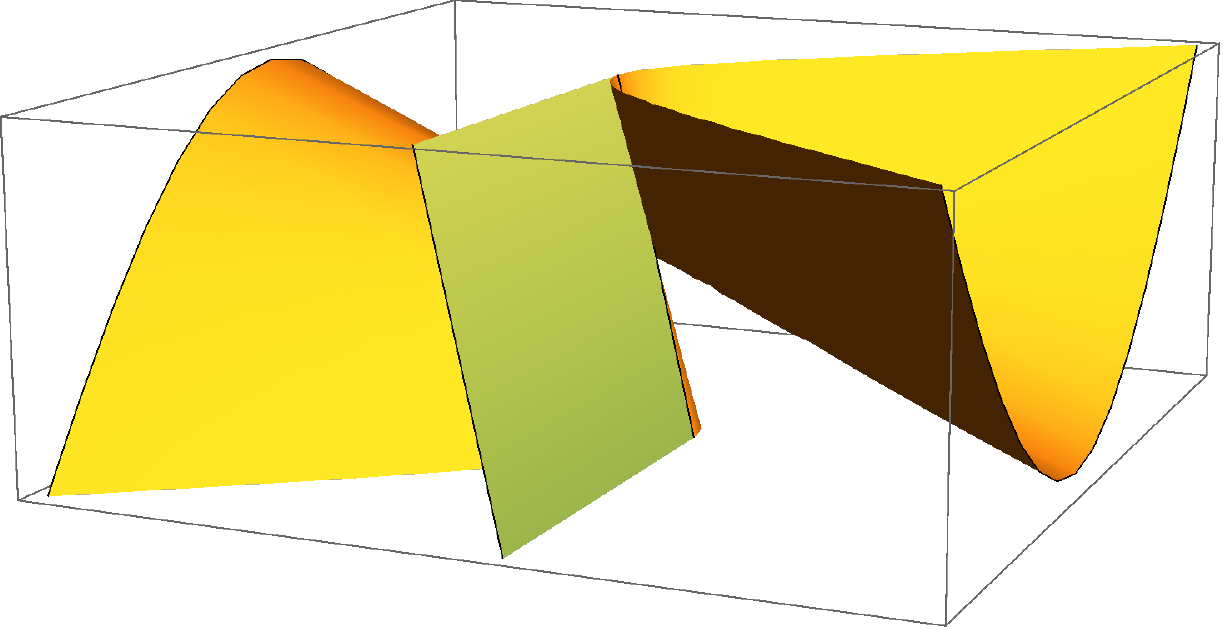}
\hspace{3mm}
\includegraphics[scale=0.3]{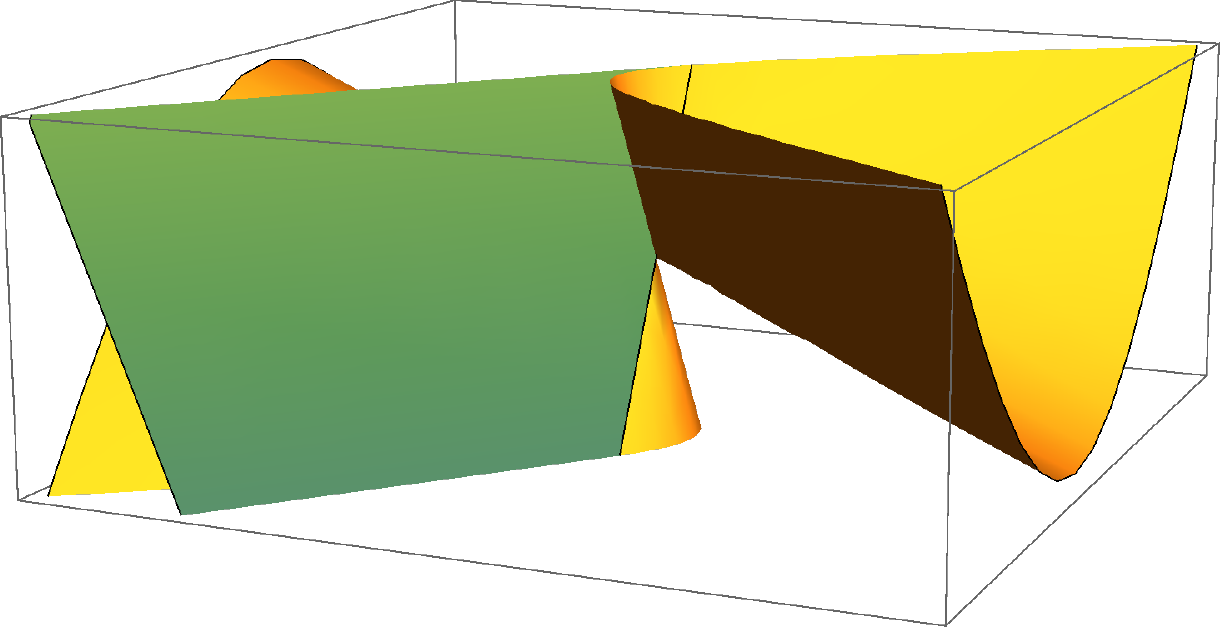}
\hspace{3mm}
\includegraphics[scale=0.3]{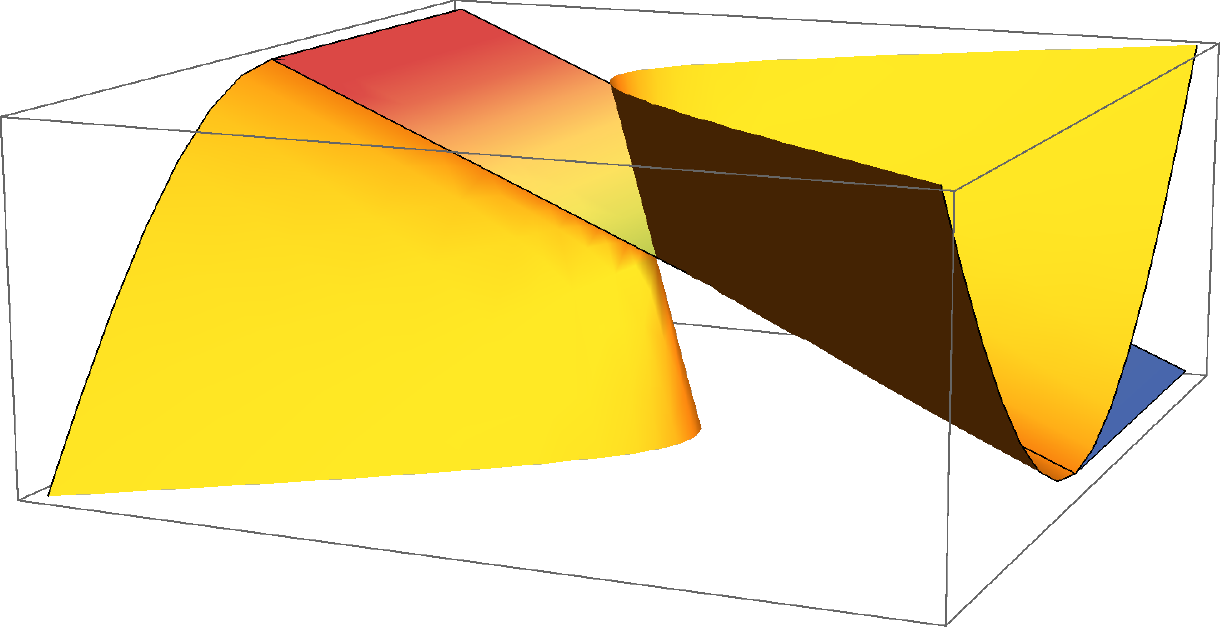}
\hspace{3mm}
\includegraphics[scale=0.3]{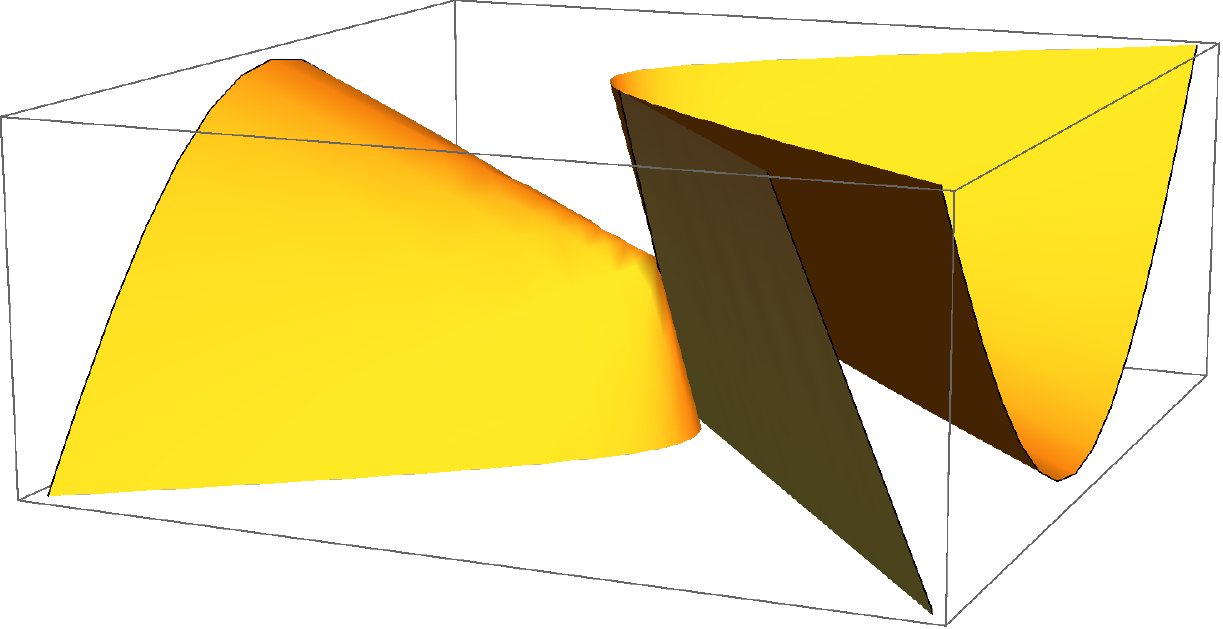}
\\[3mm]
\includegraphics[scale=0.25]{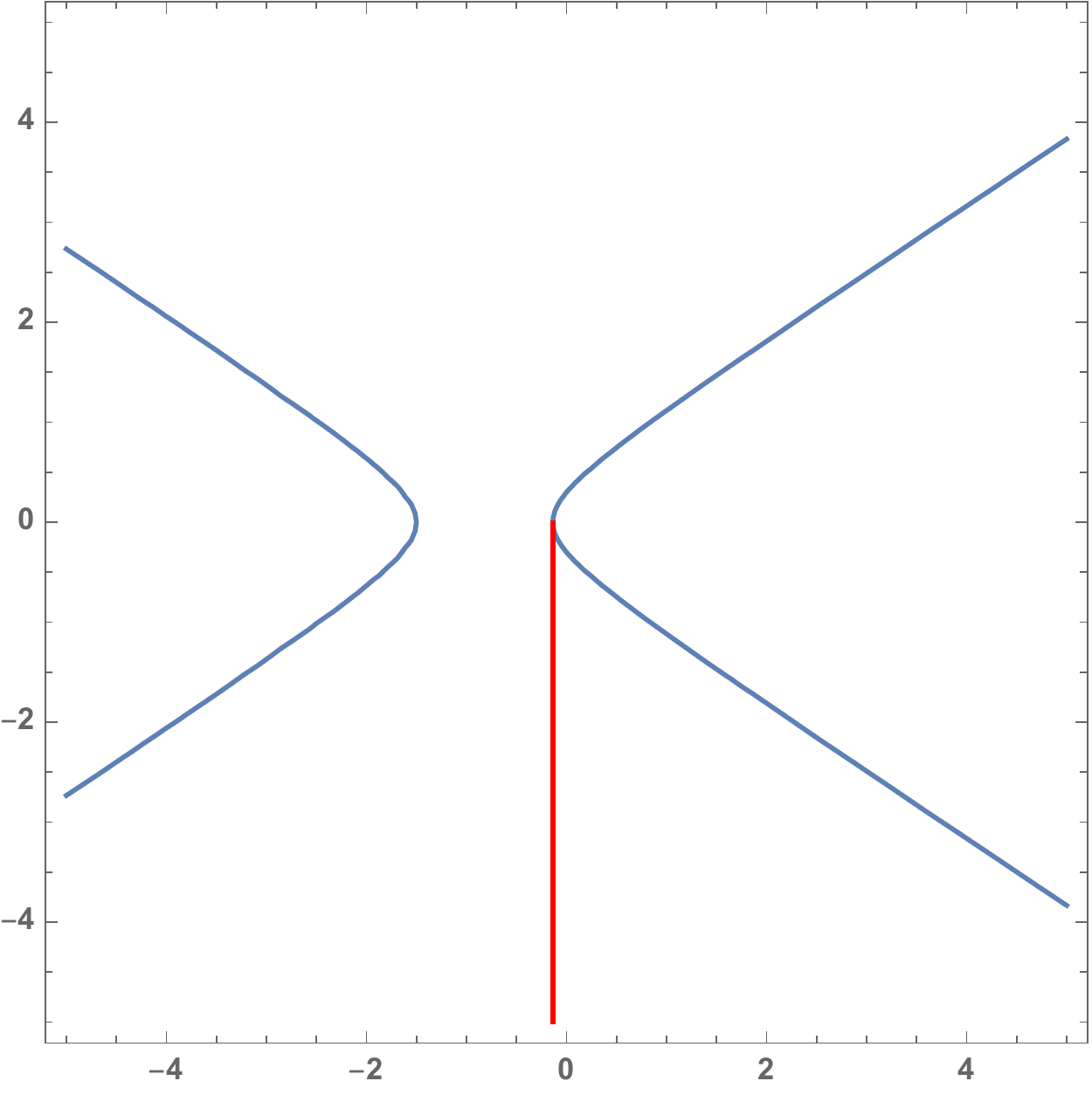}
\hspace{10mm}
\includegraphics[scale=0.25]{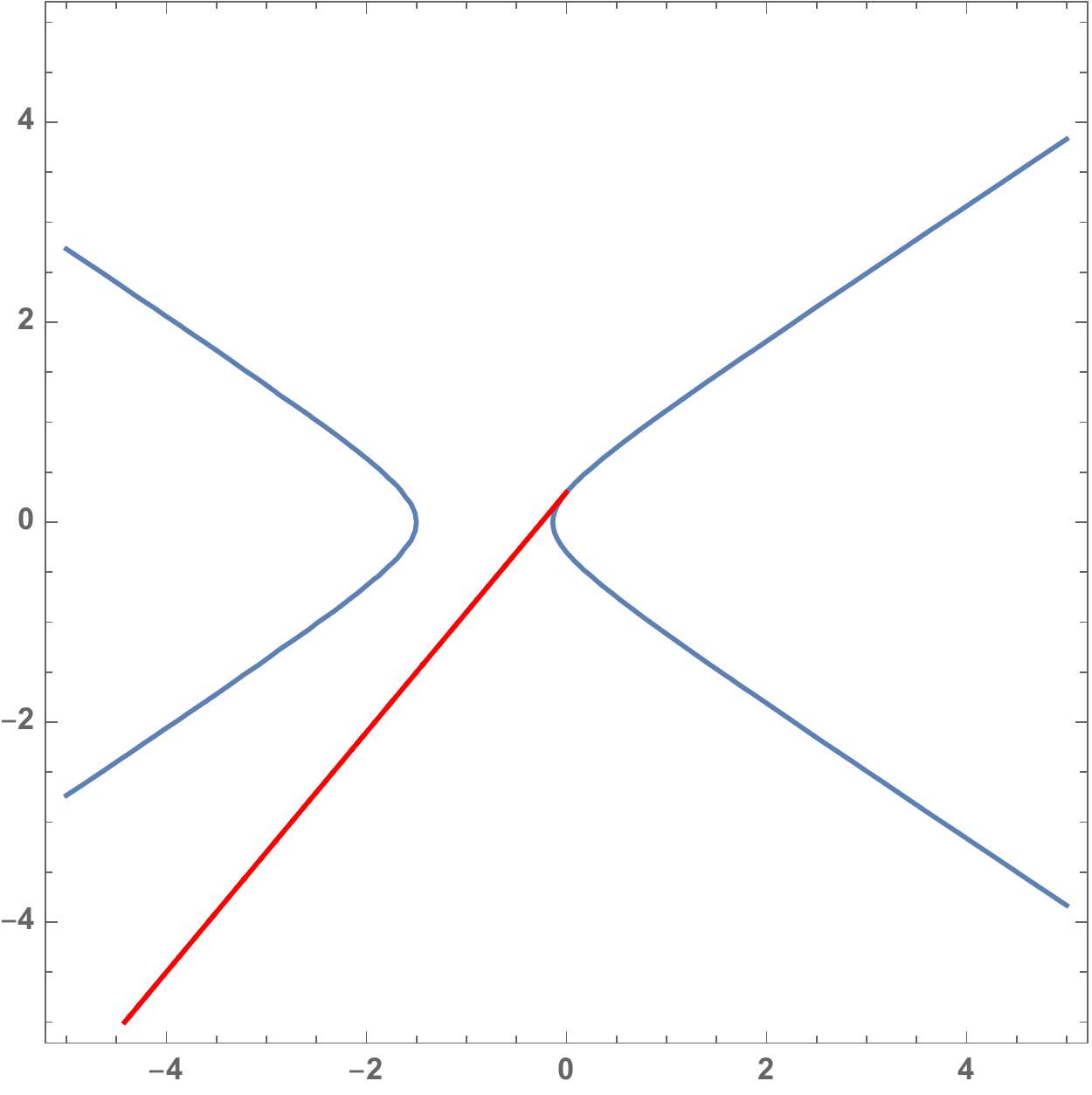}
\hspace{10mm}
\includegraphics[scale=0.25]{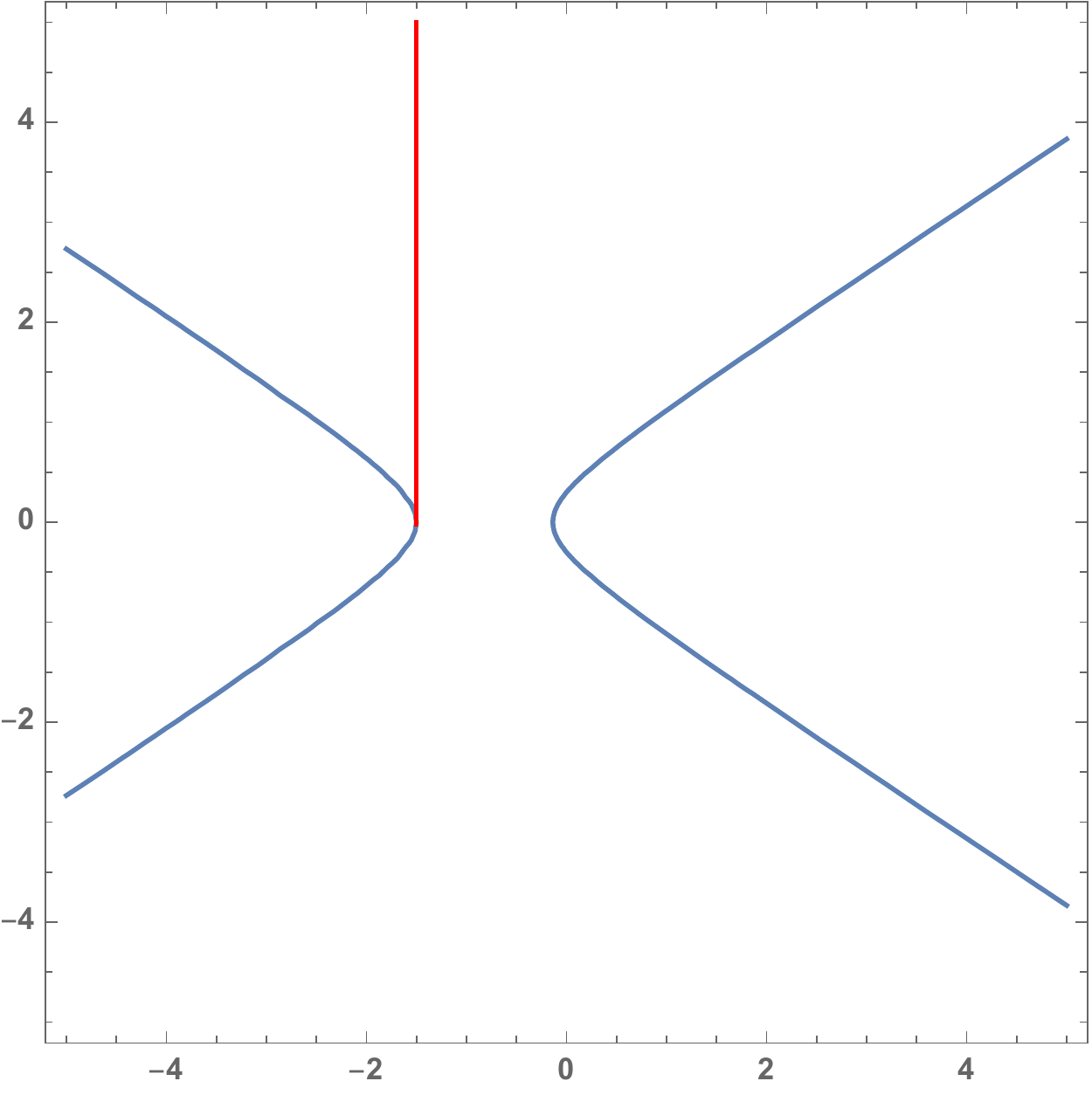}
\hspace{10mm}
\includegraphics[scale=0.25]{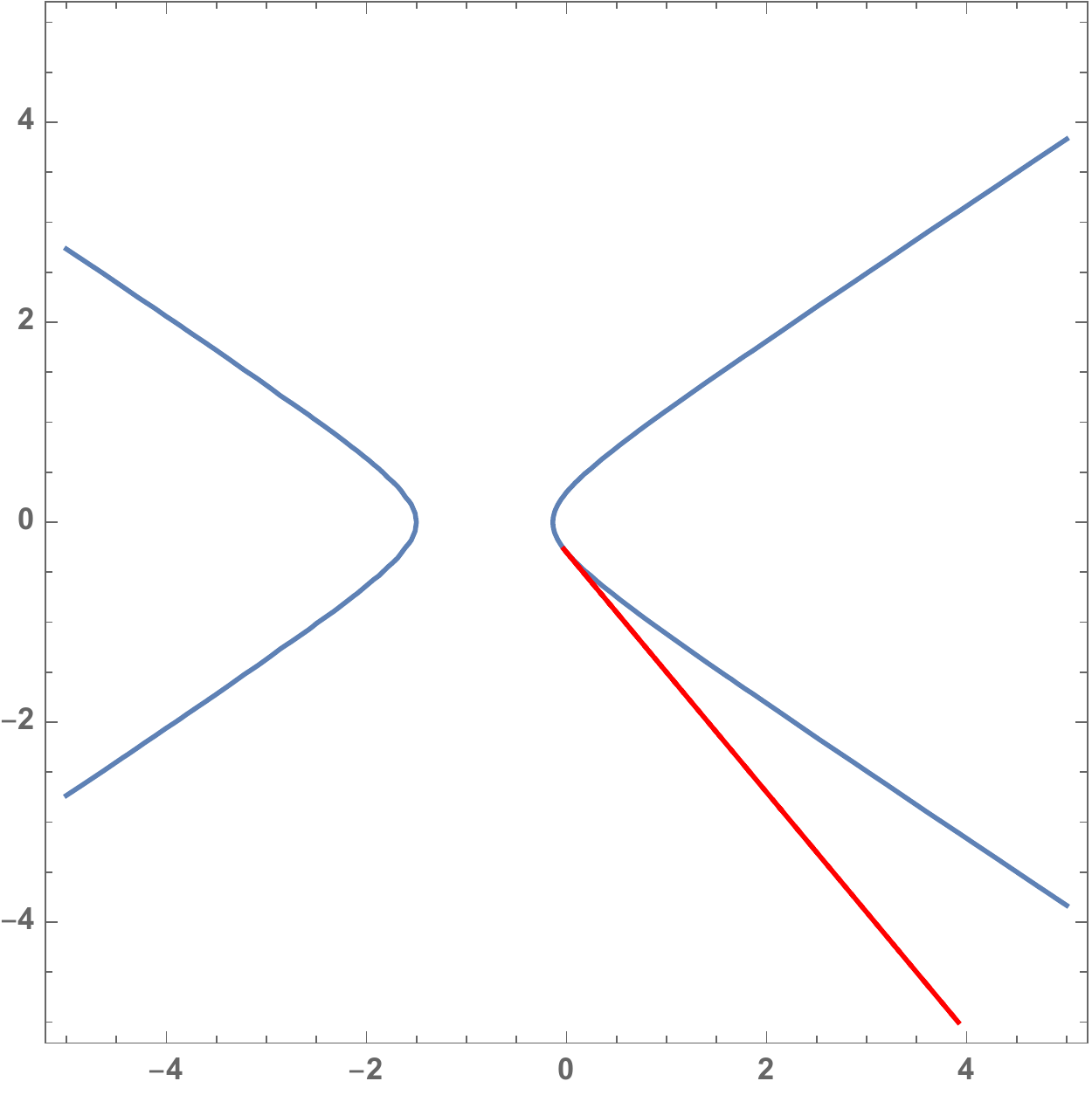}
\end{center}
\vspace{-3mm}
\caption{
Upper row: Energy dispersion $E_{\rm bulk}$ as a function of $p_1$ and $p_2$ at the slice $p_3={\rm Re}[k_3]$, and the edge dispersion $E_{\rm edge}$ given in \eqref{edgeE} with \eqref{norm}.
We chose $\alpha_2=\alpha_3=0$ with $\alpha_1=-1.2$ (Type II), and the boundary condition parameter 
$\theta=0$, $\theta=\pi/2$, $\theta=\pi$ and $\theta=(3\pi/2)$ (From left to right). 
The edge dispersion is always flat, and tangential to the bulk edge dispersion.
Lower row: Corresponding slices at $E=0.3$, shown in the $(p_1, p_2)$-plane. 
Blue curved lines are for the bulk dispersion, and 
red half lines are for the edge dispersion.
}
\label{edgefig}
\end{figure*}

\section{Edge dispersion of Type II Weyl semimetals}
\label{sec4}

The edge state should exist as a result of the topological protection, since the bulk-edge correspondence 
\cite{jackiw1976solitons,hatsugai1993chern,wen2004quantum}
works also for the Type II Weyl semimetals \cite{soluyanov2015type,mccormick2017minimal}.
The edge state is localized at the boundary because of the imaginary part 
of the momentum normal to the boundary. 
Although the bulk mode satisfies the boundary condition by taking 
an appropriate linear combination of the incoming and outgoing modes at the boundary, 
such a linear combination cannot be taken for the edge mode 
since only one of these two modes corresponds to the edge mode 
and the other is an unphysical non-normalizable mode. 
Thus, the boundary condition gives an additional condition to the momenta of the edge mode. 
 
Let us solve the Hamiltonian eigen equation $H\psi = E_{\rm edge} \psi$ for the edge states,
by imposing the most generic boundary condition \eqref{bc}.
It is quite straightforward and we show only the result here. The energy eigenvalue is
\begin{align}
E_{\rm edge} = \alpha_1 p_1 \!+ \alpha_2 p_2 \!-\! \sqrt{1\!-\!\alpha_3^2}
\left(p_1 \cos\theta \!- p_2 \sin\theta\right).
\label{edgeE}
\end{align}
The edge state wave function is
\begin{align}
\psi = \left(
\begin{array}{c}
-\sqrt{\frac{1-\alpha_3}{1+\alpha_3}} e^{i\theta}
\\
1
\end{array}
\right) \exp [i k_3 x^3]
\end{align}
with the complex momentum $k_3$,
\begin{align}
k_3 \equiv \frac{
\alpha_3 (p_1 \cos\theta - p_2 \sin\theta) -i (p_1 \sin\theta + p_2 \cos\theta) }{\sqrt{1-\alpha_3^2}}.
\end{align}
The imaginary part of $k_3$ shows the localization of the edge state at the boundary.
When the material exits in the region $x^3\geq 0$, the normalizability condition for the wave function
is $\beta \equiv {\rm Im}[k_3]>0$, which is equivalent to
\begin{align}
p_1 \sin\theta + p_2 \cos\theta < 0 .
\label{norm}
\end{align}

The edge dispersion is a straight line in the $(p_1,p_2)$ plane at the constant energy slice.
We show some of the examples of the edge and bulk dispersions in Fig.~\ref{edgefig}.
Note that the bulk dispersion is a 2-dimensional surface but 
the edge dispersion is a 1-dimensional line in the 3-dimensional momentum space of $(p_1, p_2, p_3)$. 
Fig.~\ref{edgefig} shows the plots on the slice at $p_3={\rm Re}[k_3]$ 
in the 3-dimensional momentum space, where the edge dispersion extends. 

It should be emphasized that the edge dispersion does not intersect with the bulk dispersion.
The edge dispersion always lie outside the bulk dispersion, 
since the dispersion relation in terms of the metric $g^{\mu\nu} k_\mu k_\nu = 0$ 
simply gives $g^{\mu\nu} p_\mu p_\nu = 0$ for the bulk mode while 
$g^{\mu\nu} p_\mu p_\nu = \beta^2$ for the edge mode, 
but the edge and bulk dispersion merge at the single merging point, $\beta = 0$.
(The exception is the $E=0$ slice at which the edge dispersion could overlap with
the bulk one, for some special values of $\theta$.)

One interesting observation is that the edge dispersion is always tangential to the bulk dispersion.
The next section is devoted for a proof that the edge dispersion is always tangential 
to the bulk dispersion at the merging point.


\section{Tangentiality theorem of edge and bulk dispersions}
\label{sec5}

In this section, we show that any edge dispersion is 
tangential to the bulk dispersion at the merging point. 
The statement was explicitly made by Haldane \cite{haldane2014attachment} 
for generic Weyl semimetals 
and here we provide a proof of it.
This theorem is not only for the Type II Weyl semimetals 
but applicable to any bulk and edge state 
which satisfies the definitions that we will provide below. 

We first consider the generic bulk mode. 
It is a propagating mode in the bulk of materials, and so 
the wave function of it is given in terms of the momenta
\begin{equation}
 \psi \sim e^{i p_i x^i - i E t} \ , 
\end{equation}
where $p_i$ is the spatial momenta and $E$ is the energy. 
For stable states, the energy $E$ has to be real. 
The momenta $p_i$ should also be real for the normalizability of the state. 
Thus, we assume that both $p_i$ and $E$ are real. 

The edge mode is a localized mode around the surface boundary of the material. 
It satisfies the same equation of motion 
but the momentum normal to the boundary has an imaginary part, 
\begin{equation}
 \psi \sim e^{i p_i x^i - \beta z - i E t} \ , 
\end{equation}
where $z$ is the normal direction to the boundary and 
$\beta$ is the imaginary part of the momentum in the direction. 
Thus the wave function is suppressed away from the boundary. 

The surface boundary condition needs to be imposed on the wave functions above at the boundary. 
For bulk modes, 
it can be satisfied by 
taking an appropriate superposition of the incoming mode $p_z<0$ and the outgoing mode $p_z > 0$. 
On the other hand, for the edge modes, 
these two modes would correspond to those with opposite signs of $\beta$. 
The linear combination cannot be taken due to the normalizability condition, 
and thus, the boundary condition gives an additional constraint on the momenta. 
This structure is generic, and the edge dispersion is subject to
additional constraints in general. 
The additional constraints however play no important role in the proof. 

The statement of the theorem which we prove is:
Bulk and edge modes are tangential to each other at their merging point, 
for any system which satisfies the following conditions, 
\begin{enumerate}
%
%
%
%

\item[(i)]
Bulk mode is defined as the states whose momenta are real. 

\item[(ii)]
For the edge mode, only one of the momenta has an imaginary part. 

\item[(iii)]
The energy is given by a function of momenta. 
The function is holomorphic and the form is shared for 
bulk and edge modes. 

\item[(iv)]
The energy may not be real for arbitrary complex values of momenta, 
but is real for bulk modes and edge modes. 

\end{enumerate}

And here we provide a proof.
According to the assumptions,
both the bulk and edge dispersions are given by 
subspaces of the curve 
\begin{equation}
 E = F(k_i) \ , 
\end{equation}
where $k_i$ are momenta, which are complex in general. 
The bulk dispersion is the subspace of the curve 
in which all the momenta are real, 
\begin{equation}
 E = F(p_i) \ , 
\end{equation}
where $p_i$ are real momenta. 
The edge dispersion is given in terms of the same function $F$ as 
\begin{equation}
 E = F(p_{i(\neq z)},p_z + i \beta) \ , 
\end{equation}
but the momenta satisfy additional constraints which come from the boundary condition. 
If the edge dispersion continues to $\beta=0$, it is merged into the bulk dispersion there. 

Now, it is straightforward to show that the edge dispersion is tangential to the bulk dispersion. 
The tangent space of the bulk dispersion is given by 
\begin{equation}
 0 = dE = \sum_i \frac{\partial F}{\partial p_i} dp_i \ . 
\end{equation}
On the other hand, the tangent space of the edge dispersion is expressed as 
\begin{equation}
 0 = dE = \sum_{i(\neq z)} \frac{\partial F}{\partial p_i} dp_i 
  + \frac{\partial F}{\partial p_z}\left(dp_z + i d \beta\right) \ . 
\label{tane}
\end{equation}
The Hermiticity condition for the bulk mode implies that 
all $\frac{\partial F}{\partial p_i}$ must be real since 
all real momenta $p_i$ are independent for the bulk mode. 
Then, the real and imaginary parts of \eqref{tane} give 
\begin{align}
 0 &= dE = \sum_i \frac{\partial F}{\partial p_i} dp_i \ , 
 \label{tanr}
\\
 0 &= \frac{ \partial F}{\partial p_z}  \ , 
 \label{tani}
\end{align}
respectively. 
At the merging point $\beta=0$, the first equation agrees with 
the tangent space of the bulk dispersion there. 
Therefore, the edge dispersion is tangent to the bulk dispersion at the merging point. 
The imaginary part \eqref{tani} must be satisfied on the merging point, 
for the energy of the edge mode to be real. 

Finally we emphasize again that the above proof is valid for any system, 
for example, a system on a discrete lattice, 
as long as it satisfies the conditions (i)-(iv) above, 
though in this paper we focus on the continuum limit in the Type II Weyl semimetals. 
For the case of Type II Weyl semimetals, 
it can be seen in Fig.~\ref{edgefig} that the edge dispersion 
is tangential to the bulk dispersion.

\begin{figure*}
\begin{center}
\includegraphics[scale=0.3]{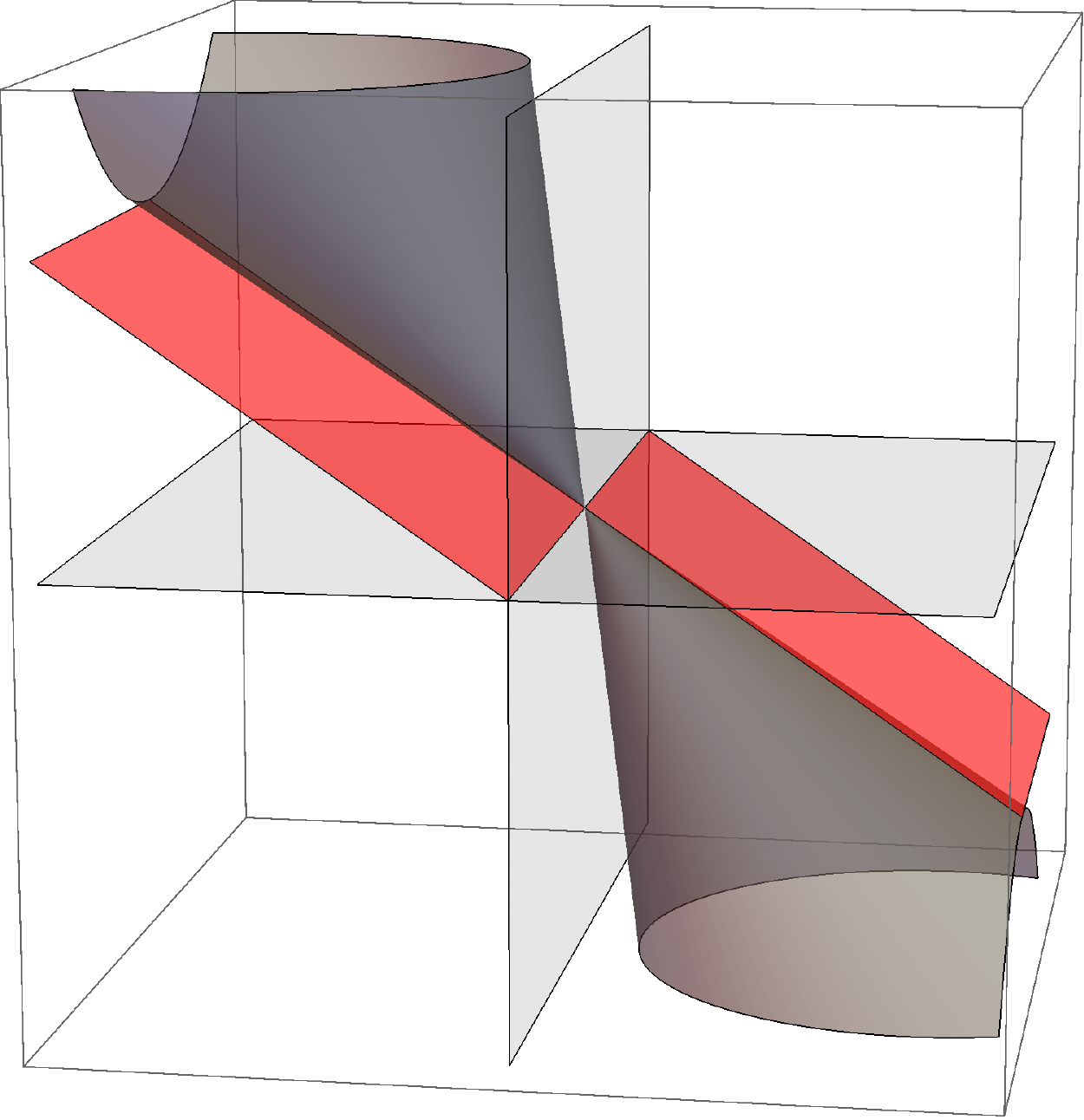}
\hspace{3mm}
\includegraphics[scale=0.3]{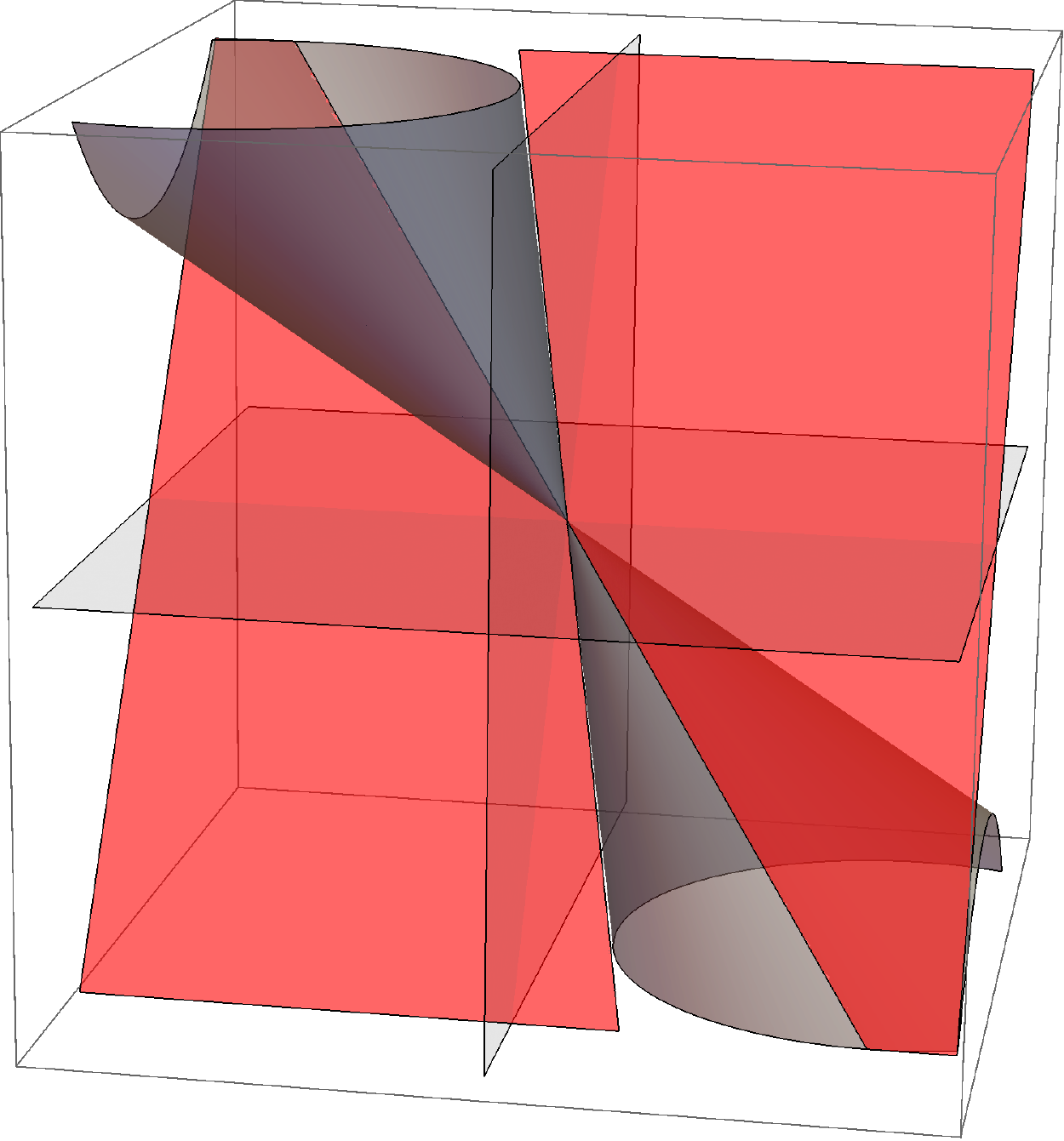}
\hspace{3mm}
\includegraphics[scale=0.3]{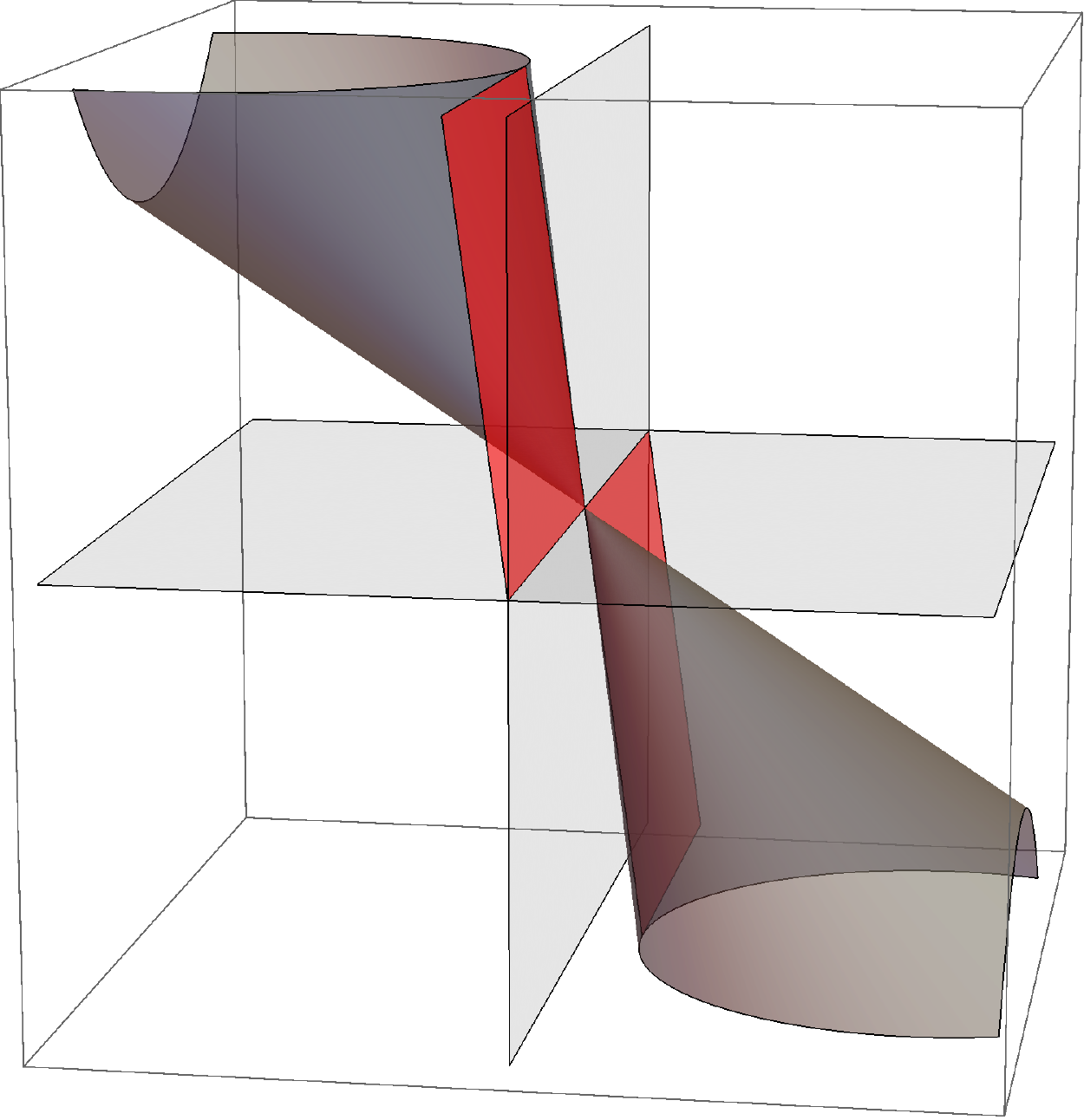}
\hspace{3mm}
\includegraphics[scale=0.3]{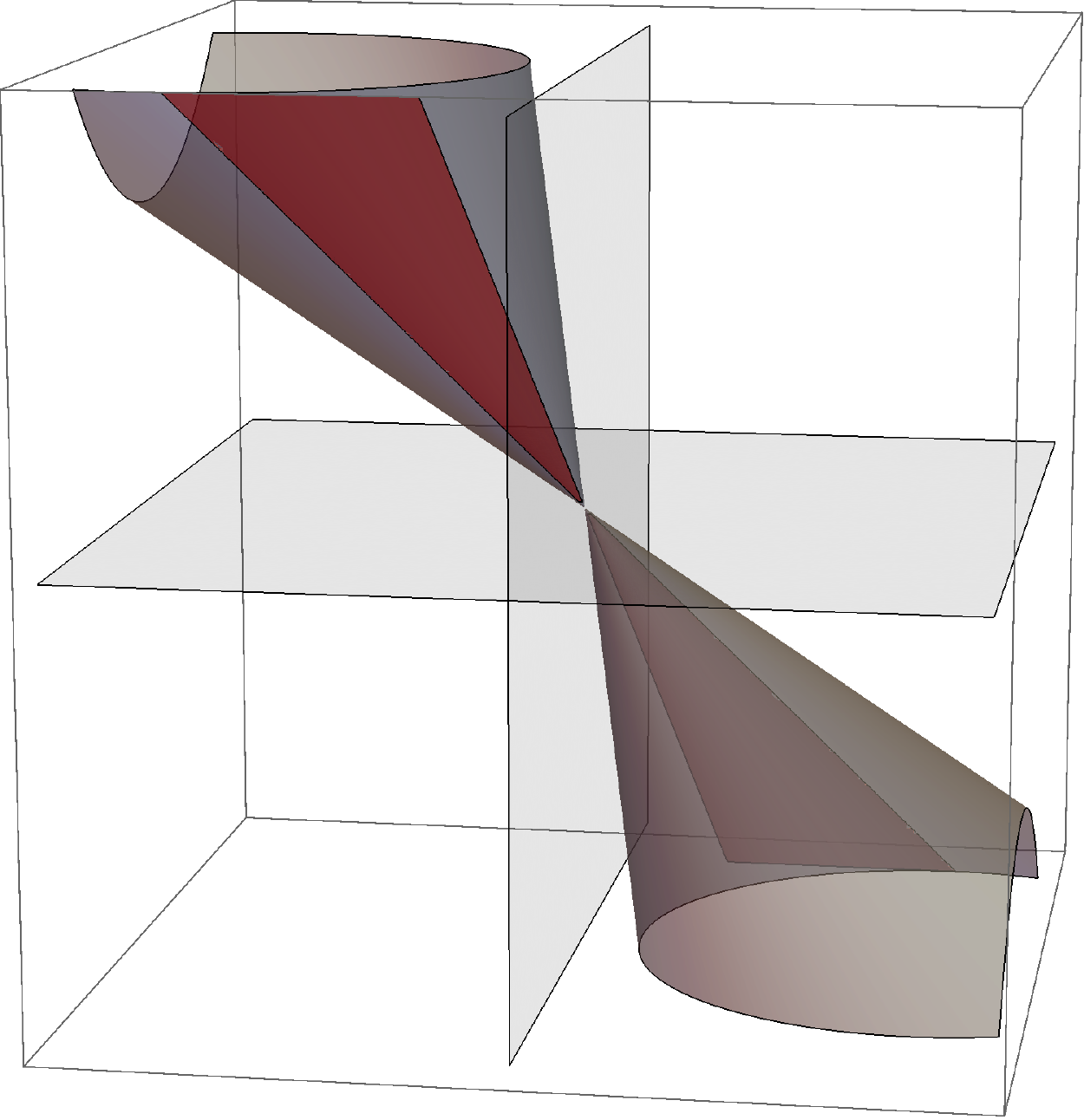}
\end{center}
\vspace{-3mm}
\caption{
Spacetime picture of the bulk (grey) and edge (red) modes.
We chose $\alpha_2=\alpha_3=0$ with $\alpha_1=-1.2$ (Type II), and the boundary condition parameter 
$\theta=0$, $\theta=\pi/2$, $\theta=\pi$ and $\theta=(3\pi/2)$ (From left to right). 
}
\label{lcfig}
\end{figure*}


\section{Escape from black holes}
\label{sec6}

Let us study the propagation direction of the edge state to see whether it can escape from
the ``black hole.'' The relation between the propagation direction $n^\mu$ and the four-momentum
$p_\mu$ is $n^\mu = g^{\mu\nu}p_\nu$. Substituting the edge dispersion \eqref{edgeE}
and $p_3={\rm Re}[k_3]$, we find
\begin{align}
& n^0=\frac{1}{\sqrt{1-\alpha_3^2}}
\left(p_1\cos\theta - p_2 \sin\theta\right), \\
& n^1=-p_1 + \frac{\alpha_1}{\sqrt{1-\alpha_3^2}}
\left(p_1\cos\theta - p_2 \sin\theta\right), \\
& n^2=-p_2 + \frac{\alpha_2}{\sqrt{1-\alpha_3^2}}
\left(p_1\cos\theta - p_2 \sin\theta\right), \\
& n^3=0.
\end{align}
Note that automatically we obtained $n^3=0$, which is consistent with the fact that
the edge mode propagates along the boundary $x^3=0$. 

The expression above applies to 
any $\alpha_1$ and $\alpha_2$. 
The Type II Weyl semimetal has $\alpha_1^2 + \alpha_2^2 > 1 - \alpha_3^2$. 
Without loss of generality, we can take $\alpha_1<-\sqrt{1-\alpha_3^2}$ and $\alpha_2=0$, 
by using the rotation in $(x^1,x^2)$-plane. 
So let us concentrate on this case. 
All the bulk modes propagate in the negative direction of $x^1$.
So, if we can find an edge mode which propagates in the positive direction of $x^1$, 
that is $n^1/n^0>0$, 
we conclude that the edge mode can escape from the black hole. 
In other words, for $(p_1,p_2)$ satisfying \eqref{norm} and $E_{\rm edge}>0$ with \eqref{edgeE},
if there exists $(p_1,p_2)$ giving $n^1/n^0>0$, the edge mode can escape from the black hole. 
As we will see below, the answer depends on the parameter $\theta$ of the boundary condition.

\begin{figure*}
\begin{center}
\includegraphics[scale=0.4]{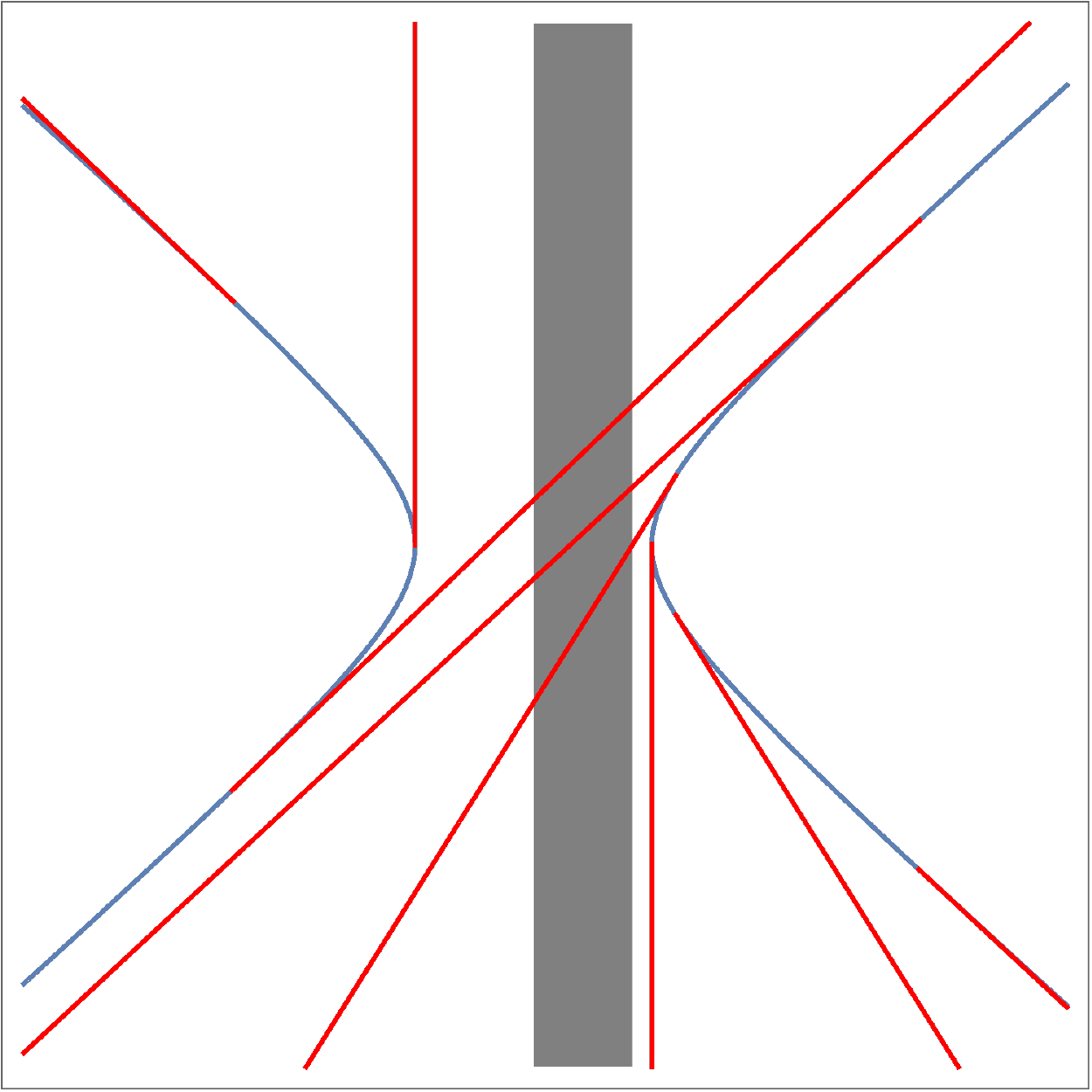}
\caption{
Edge dispersion (red line) on $E_{\rm edge} = \mathrm{const.}$ surface. 
The parameters are taken as $\alpha_1 = - 1.2$ and $\alpha_2 = \alpha_3 = 0$. 
The boundary condition parameter $\theta$ is 
$\theta = \frac{9}{8}\pi$, $\theta = \pi$, $\theta = \frac{7}{8}\pi$, 
$\theta = \frac{3}{4}\pi$, $\theta = \frac{1}{2}\pi$, $\theta = 0$, 
$\theta = \frac{3}{2}\pi$ and $\theta = \frac{5}{4}\pi$, from upper-left to lower-right. 
The shaded region \eqref{esc} lies between the hyperboloid of the bulk dispersion (blue line). 
}
\label{fig:edge2d}
\end{center}
\end{figure*}

In order to see whether the edge mode can escape from the black hole, 
it is convenient to rewrite $n^0$ and $n^1$ in terms of energy $E_{\rm edge}$; 
\begin{align}
 n^0 
 &= 
 \frac{E_{\rm edge}-\alpha_1 p_1}{1-\alpha_3^2} \ ,
 \\
 n^1 
 &= 
 \frac{\alpha_1}{1 - \alpha_3^2} E_{\rm edge} 
 + \frac{1- \alpha_1^2 - \alpha_3^2}{1 - \alpha_3^2} p_1 \ . 
\end{align}
Here, we consider only the edge modes with positive energy $E_{\rm edge} > 0$, 
and the other parameters satisfy $\alpha_1 < -\sqrt{1-\alpha_3^2}$ and $-1 < \alpha_3 < 1$. 
The sign of $n^0$ depend on given energy $E_{\rm edge}$ and momentum $p_1$ as 
\begin{align}
 n^0 &> 0 \ , &
 &\text{for}&
 p_1 &> \alpha_1^{-1} E_{\rm edge} \ , 
 \\
 n^0 &< 0 \ , &
 &\text{for}&
 p_1 &< \alpha_1^{-1} E_{\rm edge} \ , 
\end{align}
while the sign of $n^1$ flips as 
\begin{align}
 n^1 &> 0 \ , &
 &\text{for}&
 p_1 &< -\frac{\alpha_1}{1 - \alpha_1^2 - \alpha_3^2} E_{\rm edge} \ , 
 \\
 n^1 &< 0 \ , &
 &\text{for}&
 p_1 &> -\frac{\alpha_1}{1 - \alpha_1^2 - \alpha_3^2} E_{\rm edge}  \ , 
\end{align}
where both $\alpha_1^{-1} E_{\rm edge}$ and $-\frac{\alpha_1}{1 - \alpha_1^2 - \alpha_3^2}E_{\rm edge}$ are negative. 
From the conditions $E_{\rm edge} > 0$, $\alpha_1 < -\sqrt{1-\alpha_3^2}$ and $-1 < \alpha_3 < 1$,  
it is straightforward to obtain the following relation; 
\begin{equation}
 -\frac{\alpha_1}{1 - \alpha_1^2 - \alpha_3^2} E_{\rm edge} < \alpha_1^{-1} E_{\rm edge} \ . 
\end{equation}
Thus there is always a range of momentum $p_1$; 
\begin{equation}
 -\frac{\alpha_1}{1 - \alpha_1^2 - \alpha_3^2} E_{\rm edge} < p_1 < \alpha_1^{-1} E_{\rm edge} \ , 
 \label{esc}
\end{equation}
which shows $n^1/n^0 > 0$, or equivalently, a possible edge mode escaping away from the black hole. 

However, note that it does not immediately mean that there exists such an edge mode which 
can escape from the black hole.  
This edge mode needs a value of $p_1$ which is in the range \eqref{esc}, that is, 
the edge dispersion needs to allow $p_1$ to overlap with \eqref{esc}. 
This 
can be seen in Fig.~\ref{fig:edge2d}:
the edge dispersions for various $\theta$ are shown pictorially in Fig.~\ref{fig:edge2d}, 
for the case of $\alpha_1=-1.2$ and $\alpha_2=\alpha_3=0$.
If the edge dispersion (colored in red) intersects with the range \eqref{esc} (the grey region),
then that is the edge mode escaping away from the black hole.

To obtain an analytic expression for the boundary condition parameter $\theta$ to
allow such an edge mode escaping away from the black hole, we classify 
the edge dispersion by a class of ranges of $\theta$, as follows. 
\begin{itemize}
\item
For $\theta=0$, the edge dispersion is given by 
\begin{equation}
 E_{\rm edge} = \left( \alpha_1 - \sqrt{1 - \alpha_3^2} \right) p_1 \ . 
\end{equation}
Since the momentum is fixed for given $E_{\rm edge}$ 
and satisfies $0 > p_1 > \alpha_1^{-1} E_{\rm edge}$, 
the edge mode cannot escape from the black hole. 

\item
For $0 < \theta < \cos^{-1}\left( \alpha_1^{-1}\sqrt{1 - \alpha_3^2} \right) < \pi$, 
the edge dispersion has $p_2 > 0$ at the merging point $\beta = \mathrm{Im} k_3 = 0$, 
and the condition $\beta > 0$ gives the upper bound of $p_1$ but no lower bound. 
Since the edge dispersion is a straight line with $p_1 > \alpha_1^{-1} E_{\rm edge}$ 
at the merging point, 
the edge dispersion extends to the region \eqref{esc}. 
Thus the edge mode can escape from the black hole. 
\end{itemize}
\begin{itemize}
\item
For $\theta = \cos^{-1}\left( \alpha_1^{-1}\sqrt{1 - \alpha_3^2} \right) < \pi$, 
the edge mode is on the asymptote of the hyperboloid of the bulk mode. 
There is no upper or lower bound on $p_1$, and 
the edge dispersion extends to the region \eqref{esc}. 
The edge mode can escape from the black hole. 

\item
For $\cos^{-1}\left( \alpha_1^{-1}\sqrt{1 - \alpha_3^2} \right) < \theta < \pi$, 
the edge dispersion has $p_2 < 0$ at the merging point $\beta = \mathrm{Im} k_3 = 0$, 
and the condition $\beta > 0$ gives the lower bound of $p_1$ but no upper bound. 
Since the edge dispersion is a straight line with $p_1 < -\frac{\alpha_1}{1 - \alpha_1^2 - \alpha_3^2}E_{\rm edge}$ 
at the merging point, the edge dispersion extends to the region \eqref{esc}. 
Thus the edge mode can escape from the black hole. 
\end{itemize}
\begin{itemize}
\item
For $\theta=\pi$, the edge dispersion is given by 
\begin{equation}
 E_{\rm edge} = \left( \alpha_1 + \sqrt{1 - \alpha_3^2} \right) p_1 \ . 
\end{equation}
Since the momentum is fixed for given $E_{\rm edge}$ 
and satisfies $p_1 < - \frac{\alpha_1}{1 - \alpha_1^2 - \alpha_3^2}E_{\rm edge}$, 
the edge mode cannot escape from the black hole. 

\item
For $\pi < \theta < \cos^{-1}\left( \alpha_1^{-1}\sqrt{1 - \alpha_3^2} \right)$, 
the edge dispersion has $p_2 > 0$ at the merging point $\beta = \mathrm{Im} k_3 = 0$, 
and the condition $\beta > 0$ gives the upper bound of $p_1$. 
Since the edge dispersion is a straight line with $p_1 < - \frac{\alpha_1}{1 - \alpha_1^2 - \alpha_3^2}E_{\rm edge}$ 
at the merging point, which has maximum of $p_1$, 
the edge mode cannot escape from the black hole. 
\end{itemize}
\begin{itemize}
\item
For $\theta = \cos^{-1}\left( \alpha_1^{-1}\sqrt{1 - \alpha_3^2} \right) > \pi$, 
no edge mode is allowed near the Weyl point. 
The bulk dispersion is approximately given by a hyperboloid. 
The merging point of edge and bulk mode are in $p_1 \to \pm \infty$, 
and the edge dispersion extends outward from the merging point. 

\item
For $\cos^{-1}\left( \alpha_1^{-1}\sqrt{1 - \alpha_3^2} \right) < \theta < 2\pi$, 
the edge dispersion has $p_2 < 0$ at the merging point $\beta = \mathrm{Im} k_3 = 0$, 
and the condition $\beta > 0$ gives the lower bound of $p_1$. 
Since the edge dispersion is a straight line with $p_1 > \alpha_1^{-1} E_{\rm edge}$ 
at the merging point, which has minimum of $p_1$, 
the edge mode cannot escape from the black hole. 
\end{itemize}
In summary, in the convention $\alpha_1<-\sqrt{1-\alpha_3^2}$ and $\alpha_2=0$,
the edge mode can escape away from the black hole, when the
boundary condition parameter $\theta$ satisfies
\begin{align}
0<\theta<\pi \, .
\label{bhtheta}
\end{align}
This means that for a randomly chosen consistent boundary condition $\theta$, 
it may allow the edge modes propagating out of the black hole defined by the bulk mode
of the Type II Weyl semimetals. Therefore, in building a black hole analogue 
by the Type II Weyl semimetals, one needs to carefully choose the surface boundary
conditions of the material, such that the edge modes do not violate the causality
produced by the ``black hole."


Let us elaborate more on the reason for this conclusion. The effective metric \eqref{effmet}
is determined by the bulk excitations, so the light cone structure is fixed by it.
The edge modes generically propagate outside of the light cone, so {\it edge modes are
tachyonic}. With a proper choice of the boundary condition, they can even propagate
in the direction opposite to the bulk tilted light cone. Therefore the edge modes
can eventually go outside the black hole horizon.
%
%
%
%
%
%
%

\section{Summary}
\label{sec7}

In this paper, we have studied generic boundary conditions and generic edge dispersions
in Type II Weyl semimetals in the continuum and the low energy limits.
Based on the bulk dispersion argument 
\cite{volovik2017lifshitz} that Type II Weyl semimetals can be regarded 
as the inside of a black hole, we have explored possibility of having an edge mode
which can escape away from the black hole horizon. 
We have found that the generic boundary condition is parameterized by a single
rotation parameter $\theta$ ($0\leq \theta < 2\pi$) as \eqref{bc}, 
and for a part of the range of the parameter ($0<\theta<\pi$ for $\alpha_2=0$) 
there exists an edge mode escaping away from the black hole.

For a realization of the black hole by the Type II Weyl semimetals,
since any material has its surface, we need a special care about the choice of the boundary
condition. Our analysis shows that $\theta$ needs to be in the range $\pi\leq \theta\leq 2\pi$
not to violate the black hole causal structure.
A safe way 
is to choose, for example, $\theta=3\pi/2$ 
which amounts to the boundary condition
\begin{align}
\left(
1, -i \sqrt{\frac{1-\alpha_3}{1+\alpha_3}}
\right) \psi(x^3=0) = 0 \, 
\label{bc3/2}
\end{align}
for the Hamiltonian \eqref{Hamiltonian} and the spatial coordinate $x^3\geq 0$ for the material with the
surface at $x^3=0$.

In this paper we have dealt only with the continuum limit of the Type II Weyl semimetals,
because it has enabled us to study the most generic boundary conditions, which
are necessary for checking the possibility of escaping from the black hole.
The physical realization of the specific value of $\theta$ depends on discrete lattice models
of the Type II Weyl semimetal. Once the bulk discrete model is obtained, one takes the continuum limit
and extract the value of $\theta$ from the numerically observed edge mode dispersion \eqref{edgeE}, then 
one can check whether the edge mode is escaping out of the black hole or not.

The identification of the Weyl semimetals with the black hole can be extended to 
topological ``insulators." It is known that regarding one of the momenta of Weyl semimetals
to be a nonzero constant reduces the system to a topological insulator. 
The Type I Weyl semimetal with $p_i=m$ is a 2-dimensional topological insulator of class A,
and we can consider the same dimensional reduction from the Type II Weyl semimetal
to a topological ``insulator" --- which is not insulating due to the tilted light cone.
Our analysis is valid even with putting $p_2=m$. So, black hole validity can be 
checked in the same manner, with the boundary condition parameter $\theta$.

The important part of the analyses in this paper is the most generic boundary conditions
in the continuum limit. The idea of the method was used \cite{hashimoto2016topological}
to find a topological charge of the edge state, which results in the discovery of states localized 
at corners \cite{hashimoto2017edge, hashimoto_edge_of_edge} 
(which were recently called 
corner states or hinge states in higher-order topological insulators \cite{benalcazar2017quantized,schindler2018higher}).
It would be interesting to explore the edge mode contributions to the black hole interpretation of 
various deformed topological insulators, as well as Type III and Type IV Weyl semimetals
\cite{nissinen2017type}. With these deformations of the Weyl semimetal Hamiltonians, 
D-brane interpretation of the bands \cite{hashimoto2016band} may not persist, that is also an interesting issue.

Although the propagation of the bulk modes mimics that in a black hole geometry, whether the Hawking radiation
emanating from the event horizon (which is the boundary between Type I and Type II semimetals \cite{volovik2018exotic}) exists
or not is rather a subtle question, as the Hawking radiation originates in the change of the quantum vacua 
in black hole formation. It is challenging to construct a theoretical framework 
of Weyl semimetals accompanying a Hawking temperature and possible experimental set-ups \footnote{For
related studies, see Refs.~\cite{zubkov2018black, liu2018topological, huang2018black, liu2018fermionic, zubkov2018analogies, chen2019quantum}.}.

Introducing a surface boundary to the Type II Weyl semimetals in turn means slicing a black hole, 
which sounds impossible in general relativity. Black holes in brane world scenario would be the closest example
in particle physics, and we hope our condensed matter analyses may inspire also particle physics 
in the future.

\acknowledgments

We would like to thank D.~R.~Candido, H.~Katsura, M.~Koshino, M.~Kurkov, M.~Ochi, R.~Okugawa and  
A.~Zyuzin for valuable comments.
This work is supported in part by JSPS KAKENHI Grant No.~JP17H06462.

\nocite{*}

\bibliographystyle{apsrev4-1}
\bibliography{edge2}

\end{document}